# Resonance absorption of a broadband laser pulse

J.P. Palastro, J.G. Shaw, R.K. Follett, A. Colaïtis, D. Turnbull, A. Maximov,
V. Goncharov, and D.H. Froula

*University of Rochester, Laboratory for Laser Energetics, Rochester NY, USA*

**Abstract**

Broad bandwidth, infrared light sources have the potential to revolutionize inertial confinement fusion (ICF) by suppressing laser-plasma instabilities. There is, however, a tradeoff: The broad bandwidth precludes high efficiency conversion to the ultraviolet, where laser-plasma interactions are weaker. Operation in the infrared could intensify the role of resonance absorption, an effect long suspected to be the shortcoming of early ICF experiments. Here we present simulations exploring the effect of bandwidth on resonance absorption. In the linear regime, bandwidth has little effect on resonance absorption; in the nonlinear regime, bandwidth suppresses enhanced absorption resulting from the electromagnetic decay instability. These findings evince that regardless of bandwidth, an ICF implosion will confront at least linear levels of resonance absorption.

## I. Introduction

In direct drive inertial confinement fusion (ICF), an ensemble of laser pulses symmetrically illuminates a cryogenic capsule of deuterium-tritium fuel encased in a thin outer ablator [1, 2]. The illumination heats the ablator, creating a pressure that drives inward fuel compression and outward mass ejection. The compression ultimately triggers thermonuclear burn, which, under ideal conditions, sustains itself through alpha particle heating, i.e. ignition [3]. The performance of the implosion, determined by the ablation pressure, relies critically on the efficient deposition of laser pulse energy in the ablator and preserving the compressibility of the fuel. The mass ejection, however, forms a low-density plasma corona, apt for the growth of laser-plasma instabilities.

Laser-plasma instabilities inhibit energy deposition and reduce the compressibility by scattering light away from the ablator and by generating superthermal electrons, respectively [4, 5]. In cross-beam energy transfer (CBET), for instance, a mutually driven ion acoustic wave scatters incident light from one laser pulse into the outward path of an adjacent pulse [6, 7]. The decay of a laser photon into two plasmons at the quarter critical surface, or two-plasmon decay (TPD), generates hot electrons directed towards the cryogenic fuel [8, 9]. These electrons preheat the fuel, increasing the ablation pressure required for compression.

Laser-plasma instabilities can be suppressed by using broadband pulses of light [10-16]. Generally speaking, the bandwidth either detunes the interaction between waves, as in CBET [12, 13], or incoherently drives many small instabilities instead of a single coherent instability as in TPD [14, 16]. To this end, optical parametric amplifiers offer an excellent candidate for the next generation ICF driver. These amplifiers create high power, broad bandwidth infrared light that can be seeded with a variety of temporal formats [17]. There is, however, a tradeoff: The bandwidth is more than sufficient to suppress CBET and TPD, but precludes high efficiency conversion to $3^{rd}$ harmonic, where laser-plasma interactions tend to be weaker. Operation at the $1^{st}$ or $2^{nd}$ harmonic may reintroduce resonance absorption, an effect long suspected to be the shortcoming of early ICF experiments [18-21].

Resonance absorption occurs when a p-polarized electromagnetic (EM) wave, obliquely incident on a plasma density gradient, tunnels past its turning point and resonantly excites an electron plasma wave (EPW) at the critical surface [22-27]. The subsequent wave-particle interaction converts the electrostatic energy of the EPW to electron kinetic energy in the form of a superthermal tail in the electron distribution function [24, 28, 29]. The superthermal electrons, while initially directed away from the ICF capsule, reflect from the electrostatic sheath formed by their space charge separation with the ions, and impinge on the cold fuel [28, 30]. As with TPD, the electrons preheat the fuel and increase the ablation pressure required for compression.

Here we present simulations that show laser bandwidth fails to mitigate linear resonance absorption, but suppresses nonlinearly enhanced resonance absorption. Said differently: any deleterious effect to ICF resulting from hot electrons produced during linear resonance absorption cannot be remedied by bandwidth. In linear resonance absorption, in which the ponderomotive response of the ions is neglected, reduced absorption within one frequency range is offset by enhanced absorption in another. When the nonlinear ion response is included, the absorption evolves through two stages. In the first stage, the ponderomotive force of the resonantly excited EPWs steepens the ion density profile, which increases or decreases the absorption depending on the incidence angle of EM wave. In the second stage, the electromagnetic decay instability creates transverse density modulations along the critical surface. These modulations significantly enhance resonance absorption (up to 4× for the parameters of interest). Bandwidth suppresses the nonlinear modifications by reducing the ponderomotive force of the EPWs and delocalizing the instability along the density gradient, returning the absorption to linear levels.

The remainder of this manuscript is organized as follows. Section II reviews the phenomenology behind linear resonance absorption and introduces the role of bandwidth. Section III describes an extension of the Zakharov equations appropriate for modeling the linear and nonlinear resonance absorption of a broadband laser pulse. Results of LPSE (Laser Plasma Simulation Environment [31]) simulations employing these equations are presented in Section IV. Section V ends the manuscript with a summary and conclusions.

**II. Linear Resonance Absorption**

Figure (1) provides a schematic of linear resonance absorption. An incident EM wave with frequency $\omega_0$ propagates up a preformed density ramp. The density increases with the coordinate $x$. The local EM dispersion relation determines the parallel wavenumber of the wave: $k_x^{EM} = (\omega_0 / c)[\cos^2\theta - n_0(x)/n_{cr}]^{1/2}$, where $n_{cr} = m_e \varepsilon_0 \omega_0^2 / e^2$ is the critical density, $m_e$ the electron mass, $e$ the fundamental unit of charge, $n_0(x)$ the preformed electron density, and $\theta$ the incidence angle at zero density. The parallel wavenumber decreases as the phase fronts of the wave encounter higher densities. At the turning point, $n_0(x) = n_{cr}\cos^2\theta$, the wavenumber vanishes, and the EM wave splits into reflected and evanescent waves.

The evanescent wave tunnels past the turning point and reaches the critical surface, $n_0(x) = n_{cr}$. If the EM wave has a component of polarization parallel to the density gradient, it drives an oscillating charge separation by accelerating electrons back and forth across the gradient. The evanescent wave drives this oscillation at the critical surface, resonantly exciting an EPW.

The resonantly excited EPW originates near its own turning point, $n_0(x) = n_{cr}$, forcing it to propagate down the density ramp. From the local dispersion relation for the EPW, $k_x^{EPW} \approx -(\omega_0 / 3^{1/2} v_T)[1 - n_0(x)/n_{cr}]^{1/2}$, where $v_T = (k_b T_e / m_e)^{1/2}$ is the electron thermal velocity and $T_e$ the electron temperature. The phase velocity of the EPW, $v_p^{EPW} \approx \omega_0 / k_x^{EPW}$, decreases as the phase fronts encounter lower densities, causing the EPW to be increasingly Landau damped: Near the critical surface, the EPW accelerates a small number of electrons far into the tail of the velocity distribution, $|v_p^{EPW}| \gg v_T$, and undergoes little damping. At lower densities, $n_0(x) \sim 2n_{cr}/3$, the EPW accelerates electrons closer to the bulk of the distribution, $v_p^{EPW} \approx -3v_T$, and rapidly damps away. By the time the phase fronts of the EPW reach $n_0(x) \sim n_{cr}/2$, the EM energy that was converted to electrostatic energy at the critical surface will be almost entirely converted to electron kinetic energy.

The EPW accelerates electrons in the direction of its phase velocity, producing a flux of heated electrons directed down the density ramp. In an ICF implosion, the flux moves away from the capsule. However, the resulting electrostatic sheath [28, 30] can accelerate the electrons back inward. This refluxing of electrons can preheat the capsule, inhibiting the compression and ultimately limiting the yield.

Assuming a linear density ramp, $n_0 = (x/L) n_{cr}$, several analytic theories and numerical calculations have shown that the fractional absorption, the energy absorbed divided by the incident EM energy, depends on a single parameter: $q = (\omega_0 L/c)^{2/3} \sin^2 \theta$ [22]. The red curve in Fig (2) illustrates this dependence. The peak fractional absorption occurs for intermediate values of $q \sim 0.5$. Smaller values of $q$ (smaller incidence angles) reduce the parallel component of the incident electric field, which weakens the excitation of the EPW. Larger values of $q$ (larger incidence angles) extend the distance between the turning point and the critical surface. The evanescent wave must tunnel through a larger barrier, reducing its amplitude at the critical surface and weakening the excitation of the EPW.

A broadband EM pulse incident on a plasma comprises multiple frequencies, each of which undergoes resonance absorption independently. The fractional absorption for each frequency is determined solely by its value of $q_\omega = (\omega/\omega_0)^{4/3}(\omega L/c)^{2/3}\sin^2\theta$. The additional factor, $(\omega/\omega_0)^{4/3}$, ensures each frequency experiences the same, absolute density profile (i.e. the scale length, $L$, is defined in terms of the critical density of the frequency $\omega_0$). The presence of multiple frequencies has little effect on the total fractional absorption. This is illustrated by the blue curve in Fig. (2). Defining the relative bandwidth with respect to the central frequency, $\omega_0$, as $\Delta\omega/\omega_0$, the range of $q_\omega$ values is $\Delta q \approx 2(\Delta\omega/\omega_0)q$ —a weak function of the bandwidth. As we will see, the presence of nonlinearity modifies the simple linear picture dramatically.

**III. Nonlinear Resonance Absorption Model**

To model nonlinear resonance absorption, we employ an extension of the Zakharov system of equations [32]. The traditional Zakharov system describes the nonlinear coupling of high and low frequency electrostatic fluctuations. A low frequency wave equation evolves ion density perturbations and has a natural mode at the ion-acoustic frequency; a high frequency equation propagates the electric field of electron plasma waves enveloped about a local plasma frequency. A model of nonlinear resonance absorption must also describe electromagnetic wave propagation and mode conversion between electromagnetic waves and EPWs. Near the critical surface, electromagnetic waves and EPWs have similar frequencies and a single, time-enveloped equation can be used to propagate both.

We express the high frequency electric field, $\tilde{\mathbf{E}}$, as an envelope modulated by a temporal carrier: $\tilde{\mathbf{E}} = \tfrac{1}{2}\mathbf{E}(\mathbf{x},t)e^{-i\omega_0 t} + \text{c.c.}$, where $\omega_0$ is the central frequency of the incident EM wave. The envelope evolves according to

$$\left[2i\omega_0 \frac{\partial}{\partial t}\mathbf{E} + c^2 \nabla^2 \mathbf{E} - (c^2 - 3\mathrm{v}_T^2)\nabla(\nabla\cdot\mathbf{E}) + (\omega_0^2 - \omega_p^2)\mathbf{E}\right]$$
$$= \mathrm{v}_T^2 (\nabla\cdot\mathbf{E})\frac{\nabla n}{n} - i\frac{\omega_p^2}{\omega_0}\nu_{\text{ei}}\mathbf{E} + \mathbf{Q} \quad , (1)$$

where $\omega_p^2 = e^2 n / m_e \varepsilon_0$ is the plasma frequency, $n$ the total electron density, and $\nu_{ei}$ the electron-ion collision frequency as defined in Ref. [33]. The first term on the right hand side of Eq. (1) results from an electron-ion Boltzmann equilibrium in which a quasi-static electric field balances an inhomogeneous electron pressure. The second term captures the loss of electromagnetic and electrostatic energy due to inverse bremsstrahlung heating. The function $\mathbf{Q}$ accounts for the loss of electrostatic energy through Landau damping; in wavenumber space $\hat{\mathbf{Q}} = -i\omega_0 \nu_{ld} k^{-2} \mathbf{k}(\mathbf{k} \cdot \hat{\mathbf{E}})$, where the "^" denotes a quantity Fourier transformed with respect to space and $\nu_{ld} = \nu_{ld}(k)$ is the Landau damping rate calculated as in Ref. [34].

The electron density contains two contributions: an inhomogeneous, equilibrium background, representing the preformed plasma, and the low frequency density fluctuation: $n = n_0(\mathbf{x}) + n_l$. The low frequency density perturbation corresponds to ion motion. On this time scale, the electrons respond near-instantaneously, maintaining quasi-neutrality. The ponderomotive force of the mixed electromagnetic/static waves drives the low frequency density fluctuations:

$$\left[ (\partial_t + \mathbf{u} \cdot \nabla)^2 + \nu_i (\partial_t + \mathbf{u} \cdot \nabla) - c_s^2 \nabla^2 \right] n_l = \frac{Ze^2 n_0}{4 m_e m_i \omega_0^2} \nabla^2 |\mathbf{E}|^2, \quad (2)$$

where $\mathbf{u}$ is the background flow velocity, $c_s = [(ZT_e + 3T_i)/m_i]^{1/2}$ the sound speed, $Z$ the ion charge state, $T_i$ the ion temperature, $m_i$ the ion mass, and $\nu_i = \nu_i(k)$ the Landau damping of ion-acoustic waves found by solving for the low frequency root of the kinetic plasma dispersion relation.

Equations (1) and (2) compose the extended Zakharov system of equations and, in addition to resonance absorption, include effects such as Brillouin scattering, filamentation, Langmuir decay [35], the 2D radiative decay instability [36], profile steepening, and electromagnetic decay instabilities [37]. As we show below, the latter two of these play a prominent role in determining the fraction of laser pulse energy absorbed by the plasma. One can readily verify that taking the curl of Eq. (1) produces an electromagnetic wave equation with an effective source term that converts EPWs into EM waves: $2i\omega_0 \partial_t (\nabla \times \mathbf{E}) \sim (\nabla \omega_p^2) \times \mathbf{E}$. Similarly one can take the divergence of Eq. (1)

to find an electrostatic wave equation with an effective source term that converts EM waves to EPWs: $2i\omega_0 \partial_t (\nabla \cdot \mathbf{E}) \sim (\nabla \omega_p^2) \cdot \mathbf{E}$. By assuming a curl-free field, one recovers the traditional Zakharov system.

## IV. Simulations

We have implemented the extended Zakharov system of equations, Eqs. (1) and (2), into the Laser Plasma Simulation Environment (LPSE) [31]. LPSE is a computational framework for studying laser-plasma interactions relevant to inertial confinement fusion with features including: (1) non-paraxial wave propagation, (2) injection of arbitrary and realistic laser pulses, and (3) shared and distributed memory parallelization.

The LPSE simulations presented here modeled a plasma during the first picket of a direct drive, CH capsule implosion on the Omega Laser [38] at the Laboratory for Laser Energetics (see Table I for parameters). The background plasma had a linear density ramp and a constant flow from high to low density. Specifically $n_0 = (x/L)n_{cr}$ and $\mathbf{u} = -\frac{1}{2}c_s \hat{\mathbf{x}}$. In principle, the flow velocity would increase as the density decreased. However, the dynamics of interest occur close enough to the critical surface that the flow can be approximated as constant.

In each simulation, a p-polarized EM pulse in the form of a single speckle was launched up the density ramp at an angle $\theta$ with respect to the $\hat{\mathbf{x}}$-direction (see Figure (1) for a schematic). Bandwidth was applied in the form of chaotic light. Each pulse was partially coherent with a Lorentzian power spectrum, $p(\delta\omega) = \pi^{-1}[(\delta\omega)^2 + (\Delta\omega/2)^2]^{-1}(\Delta\omega/2)$, where $\Delta\omega$ is the full width at half maximum bandwidth and $\delta\omega = \omega - \omega_0$ is the shift away from the central frequency. The pulses were injected slightly inside the minimum $x$-boundary with a profile

$$\mathbf{E}_I(y,t) = \mathbf{E}_0(y) \sum_{\ell=1}^{N} [p(\delta\omega_\ell)]^{1/2} e^{i(k_{y,\ell}y - \delta\omega_\ell t + \phi_\ell)}, \quad (3)$$

where $k_{y,\ell}$ is the wavenumber in the $\hat{\mathbf{y}}$-direction accounting for the frequency shift, $\delta\omega_\ell$. The function $\mathbf{E}_0(y)$ includes a 6$^{th}$ order super-Gaussian profile perpendicular to the propagation direction with $e^{-1}$ spot size $w$ and an amplitude that ensures the correct time-averaged intensity. $N = 100$ frequencies, $\delta\omega_\ell$, and phases, $\phi_\ell$, were selected randomly from uniform distributions on the intervals $[-4\Delta\omega, 4\Delta\omega]$ and $[0, 2\pi)$ respectively. The angle of incidence, $\theta$, was varied from $0°$ to $28°$ (i.e. $q$ values ranging from $0$ to $3.4$).

The simulation domain spanned $23 \times 64.4$ $\mu$m and was divided into $1100 \times 3080$ cells with absorbing boundaries. The cell size was chosen to resolve the fine spatial structure of the low-frequency density fluctuations and to ensure convergence. The time step was chosen to satisfy the linear stability requirement of the numerical implementation of Eq. (1): $\Delta t < \omega_0 (\Delta x^2)/c^2$.

In the following, we present two cases: linear and nonlinear resonance absorption. Linear resonance absorption was simulated by excluding the low frequency density perturbation, $n_l$, from the density, $n$, when solving Eq. (1). Nonlinear resonance absorption simulations included the full density perturbation: $n_0 + n_l$. While the simulations considered the specific case of a laser pulse with a central wavelength $\lambda = 2\pi c/\omega_0 = 1.054$ $\mu$m, the results can be scaled to other wavelengths using the normalizations presented in Appendix A.

**A. Linear Resonance Absorption Results**

The red curve in Fig. 2 displays the linear fractional absorption, $f_A$, from LPSE for a monochromatic wave as a function of $q$. The absorption compares well to previous calculations [23] with two caveats: (1) LPSE includes inverse Bremsstrahlung absorption, which causes a floor in the fractional absorption of ~0.06 at all angles. (2) LPSE models an EM pulse with a finite spot size. At normal incidence, such a pulse has an electric field component parallel to the density gradient, and, in contrast to a plane wave, will undergo resonance absorption. This component of the electric field enhances the resonance absorption at small incidence angles, but is overwhelmed by the large projection of the primary polarization component onto the density gradient at large angles. The effect is readily observable in Fig. (1) as the enhancement in the fractional absorption over the collisional floor at 0 angle of incidence.

Bandwidth has almost no effect on linear resonance absorption. The blue curve in Fig (2) displays the fractional absorption averaged over 4 ps for a pulse with a relative bandwidth of $\Delta\omega/\omega_0 = 0.1$. The blue swaths indicate the 99% confidence interval calculated from the standard error in the mean. To within the confidence interval, the red and blue curves are nearly identical.

The insensitivity of linear resonance absorption to bandwidth can be demonstrated with a simple analytical model. Using heuristic arguments, Kruer derived the following expression for the fractional absorption of an EM wave with frequency $\omega$ [4]:

$$\tilde{f}_A(\omega) \simeq 2.6 q_\omega \exp(-\tfrac{4}{3} q_\omega^{3/2}). \quad (4)$$

While Eq (4) over-predicts the absorption [4], its analytic form facilitates our demonstration of the salient physics. Integrating Eq. (4) over the power spectrum of the laser pulse provides the total fractional absorption: $\tilde{f}_T = \int p(\omega) \tilde{f}_A(\omega) d\omega$. To second order in the frequency shift, one finds

$$f_T \approx \tilde{f}_A(\omega_0)\left[1 + \left(1 - 12q^{3/2} + 8q^3\right)\left(\frac{\Delta\omega}{\omega_0}\right)^2\right] \quad (5)$$

where $q = (\omega_0 L / c)^{2/3} \sin^2\theta$. The second term in Eq. (5) represents the correction to the fractional absorption due to the finite bandwidth of the pulse. For a broad relative bandwidth of $\Delta\omega/\omega_0 = 0.1$, the correction is only 2% near peak absorption, $q = 0.5$. For smaller $q$, the reduced absorption at lower frequencies almost entirely compensates the increased absorption at higher frequencies; the reverse compensation occurs for larger $q$. A large relative bandwidth only samples a small range of the fractional absorption curve, leaving the total absorption unchanged.

**B. Nonlinear Resonance Absorption Results**

The ponderomotive force of the EM wave and resonantly excited EPW drive a low frequency density perturbation that nonlinearly modifies the conversion of EM waves into EPWs: $2i\omega_0 \partial_t (\nabla \cdot \mathbf{E}) \sim (\nabla n_l) \cdot \mathbf{E}$. The result is a dynamic fractional absorption that evolves through multiple stages. Figure (3) displays the time history of the fractional absorption for a monochromatic wave. The black line shows the linear absorption for reference. At the lower intensity, $2\times10^{13}$ W/cm$^2$, the pulse undergoes reduced and enhanced absorption for $q=0.5$ ($\theta=10°$) and $q=1.8$ ($\theta=20°$), respectively. At the higher intensity, $6\times10^{13}$ W/cm$^2$, the absorption evolves through two stages. Early in time (stage 1), the absorption exhibits behavior similar to that observed at lower intensity—a reduction and enhancement for $q=0.5$ and $q=1.8$ respectively. Later in time (stage 2), the absorption increases well above the linear levels for both values of $q$.

The modifications to the absorption in stage 1 result from profile steepening [25, 39]: The ponderomotive pressure of the resonantly excited EPW pushes plasma along the density gradient, towards larger density. This is illustrated in the top left plot of Fig. (4), which shows the low frequency density perturbation during stage 1 for the $6\times10^{13}$ W/cm$^2$ pulse at $q=0.5$. The low frequency density increases from left to right across the critical surface ($x=0$). This shortens the scale length, which stretches the absorption curve to larger values of $q$ relative to the linear result. Specifically $f_A(q) \to f_A[q(L_{\text{eff}}/L)^{2/3}]$, where $L_{\text{eff}} < L$ is an effective, steepened scale length that, in general, depends on $q$ and the incident intensity. The stretching is readily apparent when comparing the linear and stage 1 absorption curves in Fig. (5).

The enhanced absorption in stage 2 results from small-scale, transverse fluctuations in the low frequency density that develop near the critical surface. These fluctuations, observable in the top right plot of Fig. (4), increase the projection of the density gradient onto the incident polarization, augmenting the mode conversion from the EM wave to the EPW. As shown by the green curve in Fig. (5), this enhances the fractional absorption well above linear levels over a range of incidence angles. For $q<1.8$, the modulations increase the absorption by at least a factor of $\sim 1.5\times$, with a maximum increase of $4\times$ at normal incidence.

The transverse density fluctuations are fueled by the EM decay instability (EDI). In EDI, an EM wave drives a parametric resonance and decays into an EPW and an ion-acoustic wave with nearly equal and opposite wavenumbers [37]. The electric field of the EM wave drives the instability by beating with the electric field of the EPW in Eq. (2) and the low frequency density perturbation in Eq. (1). The drive (and hence the instability growth) is maximized when the electric fields of all three waves are aligned. Accordingly, the EPW and ion-acoustic wave, being electrostatic, have their wavevectors aligned with the polarization of the EM wave. That is, the instability drives fluctuations transverse to the propagation direction of the incident EM wave. To satisfy phase matching, the resonance must occur close to the critical surface: the low frequency of the ion-acoustic wave forces the frequencies of the EPW and EM wave to be nearly equal.

Consistent with EDI, the right plots in Fig. (4) show that the transverse fluctuations in the low and high frequency density [ $n_h = (\varepsilon_0 / q)\nabla \cdot \mathbf{E}$ ] originate from the same location near the critical surface ( $x = 0$ ) and have a similar spatial period. Figure (6) shows the spatial spectra of the low and high frequency density fluctuations. The white circle marks the Landau cutoff, above which EPWs rapidly decay due to Landau damping, $k\lambda_d > 0.25$ with $\lambda_d = (\varepsilon_0 k_b T_e / n_{cr} e^2)^{1/2}$. The transverse fluctuations develop at the onset of stage 2, $t \sim 3.5$ ps, and are nearly mirrored in the low and high frequency densities. The fluctuations in the EPW start near their turning point, $k_x \sim 0$, and propagate down the density ramp, leading to the streaking from small to large, negative $k_x$ in the bottom right plot of Fig. (6). By the time the fluctuations reach $k\lambda_d = 0.25$, they have almost completely damped away. We note that $n_l$ and $n_h$ are real and complex quantities respectively and thus produce symmetric and asymmetric spectra.

The occurrence of EDI also explains the intensity threshold for stage 2 observed in Figs. (3) and (5). Figure (5), in particular, shows that stage 2 no longer occurs for $q > 1.8$. At larger incidence angles, the component of the EM polarization tangential to the critical surface is smaller, which weakens the drive for the instability, dropping it below threshold. Larger incidence angles also align the wavevectors of the electrostatic waves with the density gradient, limiting the interaction to a smaller region of space. Note that the instability described here is distinct from that described in Ref. [36] in two ways: (1) it is not radiative and (2) we observe no compression of the pump wave.

In principle, the EDI process itself could also increase absorption: the driven EPWs deplete the laser pulse of energy and subsequently Landau damp. This contribution to the absorption was found to be negligible. Linear simulations were initialized with the stage 2 density profiles from nonlinear simulations. The resulting fractional absorption was nearly identical to that found in the nonlinear simulations. This verified that the enhanced absorption results from augmented mode conversion at the modulated critical surface and not the EDI-generated EPWs.

In addition to the EDI, the spatial spectra in Fig. (6) exhibit several other phenomena. Linear resonance absorption manifests as the narrow streak in the high frequency density (lower left) that starts at $k_x \sim 0$ and extends to negative $k_x$. The bright, small $k_x$ features in the low frequency density (top row) correspond to profile steepening. The Langmuir decay instability, in which a pump EPW decays into a frequency down-shifted EPW and an ion-acoustic wave, also appears [35]. The resonantly excited EPW at $k_x \approx -2.5k_0$ acts as a pump for EPWs at $k_x \approx 2.5k_0$ and forward propagating ion-acoustic waves at $k_x \approx -5k_0$ —a backscattered EPW configuration consistent with maximum instability growth.

## C. Effect of Bandwidth

Bandwidth suppresses both profile steepening and the formation of transverse fluctuations in the low frequency density. The blue curve in Fig. (5) demonstrates that a relative bandwidth of $\Delta\omega/\omega_0 > 0.02$ is sufficient to return the fractional absorption to near-linear levels. The value of $\Delta\omega/\omega_0 = 0.02$ was determined by performing the scaling over relative bandwidth shown in Fig. (7) for $q = 0.5$ and $q = 1.8$.

Bandwidth suppresses profile steepening by reducing the ponderomotive force of the resonantly excited EPWs (right hand side of Eq. (2)). Each frequency within the EM pulse has a different critical surface where it resonantly excites an EPW with the corresponding frequency. This broadens the region where the EPWs originate and reduces their temporal coherence, both of which lower the electrostatic energy density. The spread in critical surface locations is $\Delta x_{cr} = (\Delta\omega/\omega_0)L$. For $\Delta\omega/\omega_0 = 0.02$ and $L = 10$ $\mu$m, $\Delta x_{cr} = 200$ nm —a significant fraction of the steepened region as evident in the top left plot of Fig. (4). The resonantly excited EPWs have a bandwidth equal to that of the EM pulse. When their coherence time, $t_c \sim 1/\Delta\omega$, is shorter than the response time of the ions, $t_i \sim 1/(k_0 c_s)$ —that is when $(\Delta\omega/\omega_0)(c/c_s) \gg 1$ —they drive the ions incoherently with a substantially lower ponderomotive force than a monochromatic wave. Said differently, for sufficient bandwidth the ions are relatively immune to the rapidly varying interference terms in the energy density of the EPWs.

Bandwidth suppresses the transverse fluctuations by spatially delocalizing the electromagnetic decay instability along the density gradient. The spatial delocalization results from two effects. First, each frequency within the EM pulse reflects from a different surface, which lowers the amplitude available to fuel the decay instability. Near the turning point, the profile of a monochromatic EM wave can be approximated as an Airy function with an effective scale length $L_A \sim 2(c^2 L/\omega^2)^{1/3}$. The profile of a broadband EM pulse is a superposition of many such functions, one for each frequency component. The frequency dependence of the turning point, $x_t = (\omega/\omega_0)^2 L\cos^2\theta$, shifts the relative location of the functions over a range $\Delta x_t \approx 2(\Delta\omega/\omega_0)L\cos^2\theta$. When $\Delta x_t \sim L_A$, the shifted Airy functions, when superposed, have substantially lower amplitude than a monochromatic wave. As an example, $\Delta x_t \sim 350$ nm and $L_A \approx 1.3$ $\mu$m for $\Delta\omega/\omega_0 = 0.02$ and $L = 10$ $\mu$m. One can show that this reduces the peak time-averaged intensity by a factor of $\sim 2$.

Second, and more importantly, each frequency within the EM pulse has its parametric resonance for the EDI at a different point along the density ramp. The bandwidth of the pulse defines a region of space where the resonance condition for each frequency is satisfied, i.e. an interaction region. At each point within the interaction region, the instability will be driven on resonance by one frequency and off resonance by every other frequency (with a detuning determined by the density scale length and distance from that frequency's resonant point). However, the effective amplitude driving the instability at each location, comprised of the resonant and off resonant contributions, is considerably lower than the peak amplitude. Compare, for instance, to a monochromatic wave, which resonantly drives the instability with its peak amplitude at a single point. As a transverse instability, the increased extent of the interaction region due to bandwidth does not compensate the reduced, effective amplitude at each location (c.f. Ref. [40]).

This second form of suppression is effective when the spread in critical surface locations is comparable to the spatial extent of the monochromatic instability, i.e. $\Delta x_{cr} \sim \Delta x_0$. From the gain bandwidth, the extent of the monochromatic instability is given by $\Delta x_0 = 2L(\gamma_{EDI}/\omega_0)$, where $\gamma_{EDI}$ is the homogeneous growth rate. At normal incidence, $\gamma_{EDI} = \frac{1}{4}(Zm_e/m_i)^{1/2}(c^2 k_y \omega_0 / c_s)^{1/2}(v_{osc}/c)$, where $v_{osc} = eE_0/m_e\omega_0$ and $E_0$ is the amplitude of the electromagnetic wave [37]. For $\Delta\omega/\omega_0 = 0.02$ and the parameters considered here, $\Delta x_{cr}/\Delta x_0 = \Delta\omega/2\gamma_{EDI} \sim 2$, consistent with the simulations.

Note that for small relative bandwidths, $\Delta\omega/\omega_0 \approx 0.001$, Fig. (7) exhibits an enhancement in the fractional absorption. Near threshold, small bandwidths introduce intensity fluctuations, which can increase the exponentiation of the instability, but do not provide a degree of incoherence sufficient for mitigation.

**V. Summary and Conclusions**

We have examined the effect of bandwidth on resonance absorption. In traditional, linear resonance absorption, bandwidth has no effect on the fractional absorption. For small angles of incidence, the suppressed absorption at lower frequencies is almost entirely offset by the enhanced absorption at higher frequencies. The reverse compensation occurs for large angles of incidence.

Bandwidth can, however, modify absorption when a laser pulse and resonantly excited EPW nonlinearly modify the ion density. At early times, the ponderomotive force of the resonantly excited EPW steepens the density profile leading to a shorter scale length. The steepening can either enhance or reduce the absorption depending on the incidence angle of the EM wave. Later in time, the nonlinear coupling between the EM wave, EPW, and low frequency density drives an electromagnetic decay instability that significantly enhances the fractional absorption of laser light (up to $4\times$). The instability modulates the critical surface, which increases the projection of the density gradient onto the polarization of the EM wave and thus augments the conversion of EM energy to electrostatic energy.

Bandwidth disrupts these processes by (1) spreading the region and frequencies over which EPWs are resonantly excited and by (2) spatially delocalizing the electromagnetic decay instability along the density gradient. The spatial delocalization results from two effects. First, each frequency component of the incident EM wave has a different turning point, which reduces the peak amplitude. Second, each frequency resonantly drives the instability at a single density, while elsewhere contributing a weakened, detuned drive. The effective amplitude at each point, comprising the resonant and detuned contributions, is considerably less than the peak amplitude, which drives the monochromatic instability. For parameters relevant to the initial picket in a direct drive implosion on the Omega laser, a relative bandwidth of $\Delta\omega/\omega_0 > 0.02$ was sufficient to suppress the enhanced absorption.

These results will inform the development of the next generation inertial confinement fusion (ICF) driver, which will require large bandwidth to suppress instabilities such as cross beam energy transfer, two-plasmon decay, and stimulated Raman scattering. To this end, optical parametric amplifiers present an excellent candidate. The amplifiers provide enormous bandwidth, $\Delta\omega/\omega_0 \sim 0.1$, at a wavelength of $\sim 1$ $\mu$m. While the light can be frequency doubled, the efficiency of frequency tripling, due to the large-bandwidth, is prohibitively low with current technology. In an ICF implosion driven at a wavelength of 1 $\mu$m or 0.5 $\mu$m, resonance absorption will play a greater role in the laser-target coupling; such implosions will still be susceptible to hot electrons generated by linear resonance absorption.

**Acknowledgements**

The authors would like to thank C. Dorrer, D. Cao, I. Igumenshchev, S. Hu, D. Haberberger, J. L. Shaw, D. Edgell, A. Howard, L. Nguyen, and D. Ramsey for useful discussions.

This work was supported by the Department of Energy under Cooperative Agreement No. DE-NA0001944, the University of Rochester, and the New York State Energy Research and Development Authority.

**Appendix: Scaling to Other Frequencies**

With a judicious choice of variable normalizations, a solution of Eqs. (1) and (2) for one set of physical parameters can be extended to a wide range of parameters. We normalize time, space, electric field, and density by $3\zeta/2\omega_0$, $3\zeta^{1/2}v_T/2\omega_0$, $(\eta/3\zeta)^{1/2}(4m_e\omega_0 v_T/e)$, and $4n_{cr}/3\zeta$, respectively, where $\zeta = m_i/Zm_e\eta$ and $\eta = 1+3T_i/ZT_e$—the usual normalizations for the Zakharov equations [41] with the plasma frequency replaced by the laser frequency and the background density set to the critical density. Equations (1) and (2) become

$$\left[i\frac{\partial}{\partial t}\mathbf{E} + c_T^2\nabla^2\mathbf{E} - (c_T^2-1)\nabla(\nabla\cdot\mathbf{E}) + \left(\frac{3\zeta}{4}-n\right)\mathbf{E}\right] = \frac{1}{3}(\nabla\cdot\mathbf{E})\frac{\nabla n}{n} - \frac{2}{3\zeta}in\nu_{ei}\mathbf{E} + \mathbf{Q} \quad (A1)$$

$$\left[(\partial_t + \mathbf{u}\cdot\nabla)^2 + \nu_i(\partial_t + \mathbf{u}\cdot\nabla) - \nabla^2\right]n_l = \nabla^2|\mathbf{E}|^2, \quad (A2)$$

where the use of the normalized, dimensionless variables is implied, $c_T^2 = c^2/3v_T^2$, and $\hat{\mathbf{Q}} = -\tfrac{1}{2}i\nu_{ld}k^{-2}\mathbf{k}(\mathbf{k}\cdot\hat{\mathbf{E}})$. Conveniently, the laser frequency appears nowhere in Eqs. (A1) and (A2). Thus given a solution at one frequency (or wavelength), one can find a solution at any other frequency by appropriately scaling time, space, electric field, and density.

| Plasma Parameters | Value |
|---|---|
| $ZT_e/T_i$ | 3.1 |
| $Zm_e/m_i$ | $3\times10^{-4}$ |
| L (μm) | 10 |
| u ($c_s$) | 0.5 |
| $v_i$ ($kc_s$) | 0.17 |
| **Pulse Parameters** | **Value** |
| λ (μm) | 1.054 |
| w (μm) | 6 |

Table 1. Parameters for simulations

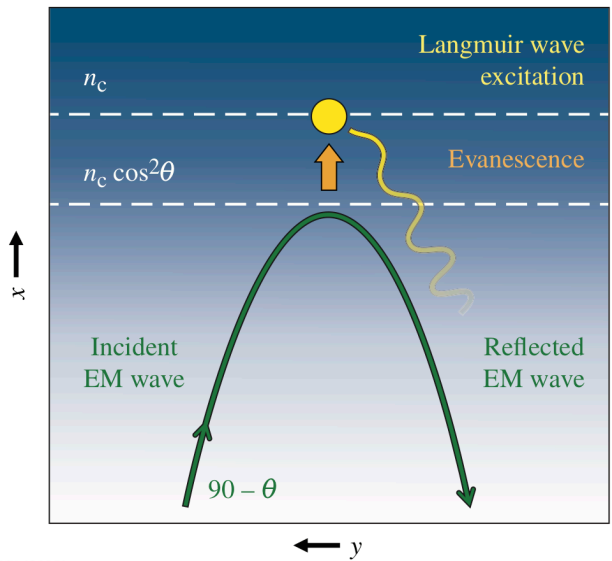

TC14320J1

Figure 1. A schematic of resonance absorption. A p-polarized electromagnetic wave reflects at its turning point. The evanescent field of the electromagnetic wave resonantly excites an electron plasma wave that propagates down the density ramp. The electron plasma wave-particle interaction converts the electrostatic energy to electron kinetic energy.

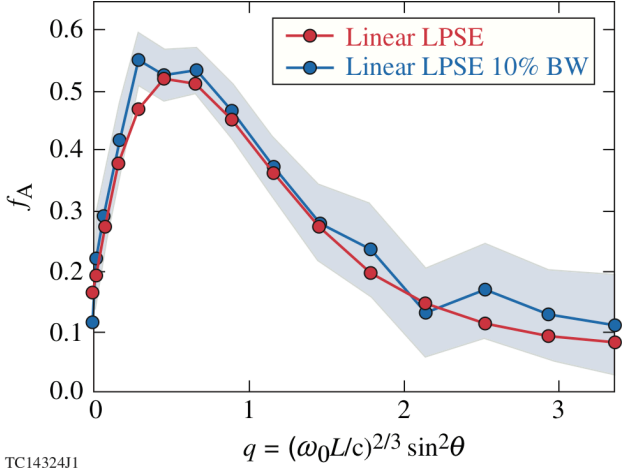

Figure 2. Fractional absorption of laser pulse energy as a function of $q$ for a monochromatic wave, red, and a broadband pulse with a relative bandwidth $\Delta\omega/\omega_0 = 0.1$, blue. For the broadband pulse, the dots were found by averaging the absorption over 4 ps. The blue swath represents the 99% confidence interval calculated from the standard error in the mean.

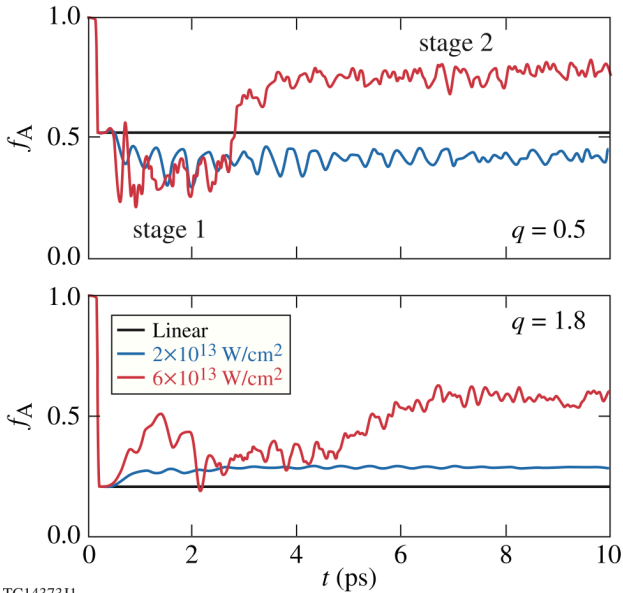

Figure 3. Fractional absorption as a function of time for a monochromatic wave undergoing nonlinear resonance absorption. The lower intensity, $2\times10^{13}$ W/cm$^2$, is below threshold for the electromagnetic decay instability, while the higher intensity, $6\times10^{13}$ W/cm$^2$, is above. The value $q = 0.5$ in the top figure was chosen to coincide

with the peak in the linear fractional absorption. The bottom figure shows a larger incidence angle, $q = 1.8$. For reference, the black line delineates the linear fractional absorption at the two $q$ values.

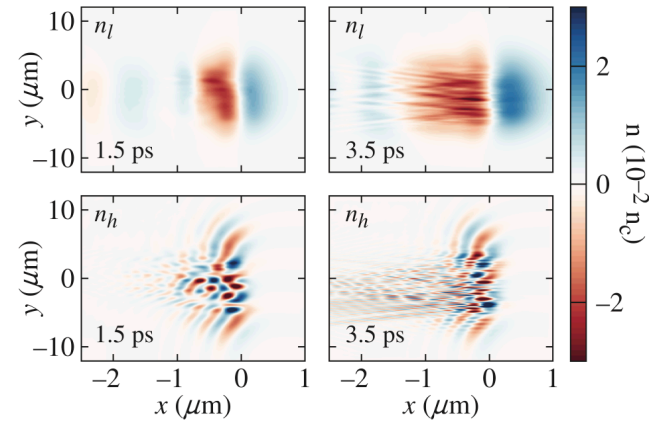

Figure 4. Spatial profiles of the low (top) and high (bottom) frequency density perturbations occurring during the absorption of a monochromatic EW wave with an incident intensity of $6 \times 10^{13}$ W/cm$^2$ and $q = 0.5$. The unperturbed critical surface is at $x = 0$. The left and right plots show the profiles in the 1$^{st}$ and 2$^{nd}$ absorption stages respectively.

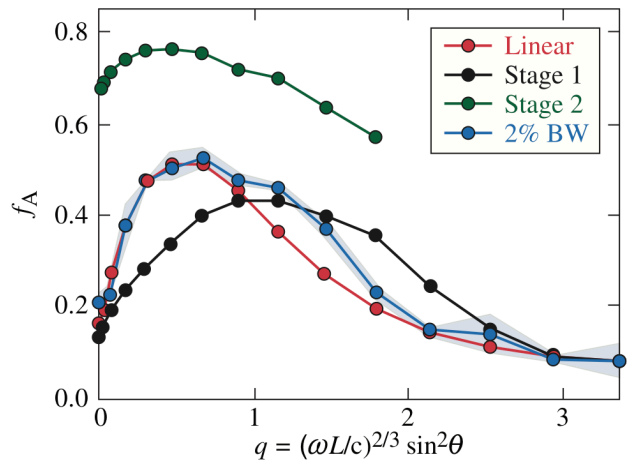

Figure 5. Fractional absorption as a function of $q$ during the 1$^{st}$ (black) and 2$^{nd}$ (green) stages of nonlinear r absorption for a monochromatic wave with an incident intensity of $6 \times 10^{13}$ W/cm$^2$. The fractional absorption returns to near-linear levels (red) when the pulse has a relative bandwidth of $\Delta\omega/\omega_0 = 0.02$ (blue). The blue dots represent the mean of the time-averaged absorption from 3 simulations. The blue swath represents the standard error in the mean of the 3 simulations.

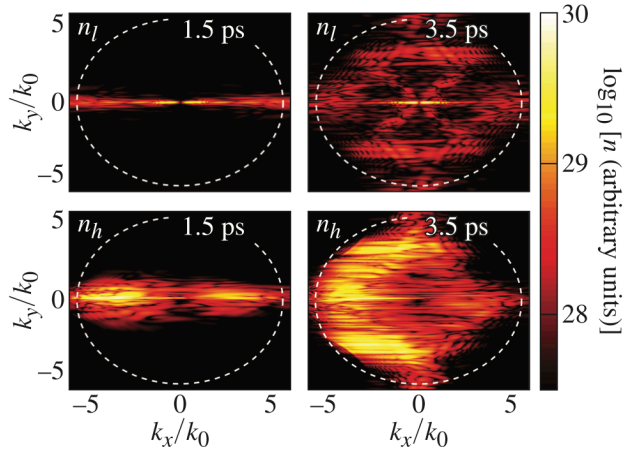

Figure 6. Spatial spectra of the high (top) and low (bottom) frequency density perturbations corresponding to the profiles displayed in Fig. (6). The white, dashed circle indicates the Landau cutoff.

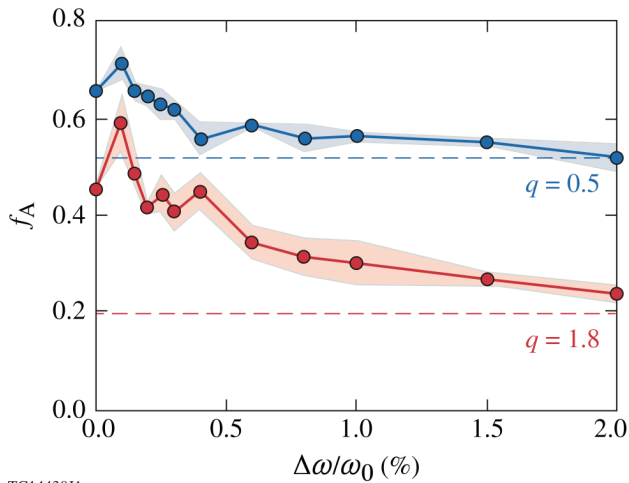

Figure 7. Fractional absorption as a function of $\Delta\omega/\omega_0$ (shown as a percentage) for $q=0.5$ (blue) and $q=1.8$ (red). The dots represent the mean of the time-averaged absorption from 3 simulations. The swath represents the standard error in the mean of the 3 simulations.